\documentclass[reqno,a4paper,11pt]{amsart}

\usepackage{amsmath,amstext,amsfonts,amsbsy,eucal,amssymb}
\usepackage[latin1]{inputenc}

\numberwithin{equation}{section}

\newtheorem{theorem}{Theorem}[section]
\newtheorem{lemma}[theorem]{Lemma}

\theoremstyle{definition}
\newtheorem{definition}[theorem]{Definition}

\newtheorem{remark}[theorem]{Remark}

\begin{document}

\parskip 4pt
\baselineskip 16pt


\title[On a certain system of discrete equations with the Laurent property]
{On a certain system of discrete equations with the Laurent property}

\author{Andrei K. Svinin}
\email{svinin@icc.ru}
\address{Matrosov Institute for System Dynamics and Control Theory, Siberian Branch of Russian Academy of Sciences, PO Box 292, 664033 Irkutsk, Russia}
\date{\today}

\begin{abstract}
We consider a system of two bilinear discrete equations with four parameters. It defines a certain discrete dynamical system. Each of its solutions is represented as a pair of sequences. It is proved that for arbitrary initial data all terms of both sequences are Laurent polynomials in the initial data with integer coefficients. This property is called the Laurent property. The proof relies on an established connection between the system and the Somos-5 equation. It is shown that both components of the solution satisfy the same equation with the same parameters. To prove the main theorem, the strong Laurent property for the Somos-5 equation is used, which makes it possible to control negative powers of the initial data. An analogue of this property is obtained directly for the system under study. As corollaries, sufficient conditions for the integrality of solutions are found. An example of a pair of integer sequences illustrating the theorem on the strong Laurent property is given.
\end{abstract}

\maketitle

\section*{Introduction}

Recently, quite a lot of research has been devoted to polynomial recurrence relations which, under certain conditions, can generate integer sequences \cite{Ustinov1,Gale,Malouf}. Interest in them is explained by a wide range of applications in algebraic combinatorics, number theory, cryptography and theoretical physics. A characteristic property of such equations is the Laurent property \cite{Alman, Fomin}: each term of the generated sequence is expressed as a Laurent polynomial in the initial data with integer coefficients.

Many examples of equations with the Laurent property of the form
\(
x_nx_{n+N}=L(x_{n+1},\ldots, x_{n+N-1}),
\)
where \(L\) is a polynomial in the indicated variables, are known in the literature, however, of direct interest to us is the bilinear equation
\begin{equation}
x_n x_{n+5}=\alpha_0 x_{n+2} x_{n+3}+\alpha_1 x_{n+1} x_{n+4}.
\label{408}
\end{equation}
Here it is assumed that \((\alpha_0, \alpha_1)\) is a pair of arbitrary parameters.
Relation \eqref{408} is known in the literature as the Somos-5 equation.

In this paper we consider a system of two bilinear discrete equations
\begin{equation}
\begin{array}{l}
x_{n} y_{n+3} = a\, x_{n+1} y_{n+2} + b\, x_{n+2} y_{n+1}, \\[0.3cm]
y_{n} x_{n+3} = d\, y_{n+1} x_{n+2} + c\, y_{n+2} x_{n+1},
\end{array}
\label{45777}
\end{equation}
where \((a,b,c,d)\in \mathbb C^4\) are arbitrary parameters, and \((x_n,y_n)_{n\in\mathbb Z}\) is a pair of sought sequences. Any solution of this system is determined by the initial data
\((x_0,x_1,x_2,y_0,y_1,y_2)\in \mathbb C^6\) and the choice of parameter values.
\begin{remark}
The use of the same notation \((x_n)\) for the sought sequence both in \eqref{408} and in \eqref{45777} is not accidental. We do this for convenience, since in the next section it will be shown that any solution \((x_n, y_n)\) of system \eqref{45777} generates a pair of solutions of the Somos-5 equation, defined as relation \eqref{408}, with suitable values of the parameters \((\alpha_0, \alpha_1)\) depending on the data
\begin{equation}
(x_0, x_1, x_2, y_0, y_1, y_2; a, b, c, d)\in \mathbb C^{10}.
\label{7666543}
\end{equation}
\end{remark}

\begin{definition}
A solution of system \eqref{45777} will be called non-singular if
\(x_n\neq0\) and \(y_n\neq0\) for all \(n\in\mathbb Z\).
\end{definition}

Interest in system \eqref{45777} is explained by the fact that, as we will show below, under certain conditions it generates a pair of integer sequences.
\begin{remark}
Systems of bilinear recurrence relations of the form \eqref{45777} were considered earlier \cite{Avdeeva,Bykovskii}. In particular, in \cite{Bykovskii} a system of equations similar to ours was studied
\[
\begin{aligned}
x_n y_{n+4}&=a x_{n+1} y_{n+3}+b x_{n+2} y_{n+2}, \\
y_n x_{n+4}&=d y_{n+1} x_{n+3}+c y_{n+2} x_{n+2}.
\end{aligned}
\]
For it, the Laurent property was proved and a general solution was constructed in terms of the Weierstrass function.
\end{remark}
\begin{theorem}
\label{6754387}
Let \((x_n,y_n)\) be a non-singular solution of system \((\ref{45777})\).
Then for all \(n\in\mathbb Z\)
\[
x_n,\;y_n\in \mathbb Z \bigl[
x_0^{\pm1},x_1^{\pm1},x_2^{\pm1},
y_0^{\pm1},y_1^{\pm1},y_2^{\pm1}; a,b,c,d
\bigr].
\]
\end{theorem}
In other words, each term of the sequences \((x_n)\) and \((y_n)\)
is a Laurent polynomial in the initial data \((x_0,x_1,x_2,y_0,y_1,y_2)\) with coefficients from \(\mathbb Z[a,b,c,d]\). In this case we say that the system possesses the Laurent property with respect to the initial data.

Practically, the meaning of this theorem is that it gives a sufficient criterion for constructing integer sequences. Indeed, let \(x_j, y_j=\pm1\) for
\(j=0,1,2\). Then by virtue of Theorem \ref{6754387} under this condition we obtain
\(
x_n,\; y_n\in\mathbb{Z}[a, b, c, d].
\)
If we now choose some integer values for the parameters \(a,b,c,d\), then \(x_n,y_n\in\mathbb{Z}\). Note that the condition given above is sufficient but not necessary. Integer sequences can also arise for other initial data and parameters.

As for the Somos-5 equation, the following well-known statement can be formulated for it.
\begin{theorem}
\label{67543}
By virtue of (\ref{408}) the following holds:
\begin{equation}
x_n\in\mathbb{Z}[x_0^{\pm 1}, x_1^{\pm 1}, x_2^{\pm 1},
x_3^{\pm 1}, x_4^{\pm 1}; \alpha_0, \alpha_1]\;\; \forall n\in\mathbb{Z}.
\label{455543377}
\end{equation}
\end{theorem}
As we have already said, system \eqref{45777} generates two sequences satisfying the same Somos-5 equation.
Therefore Theorem \ref{6754387} could in principle be a consequence of Theorem \ref{67543}. This is indeed so, however it turns out that
precisely in such a formulation Theorem \ref{67543} cannot be directly applied due to certain technical difficulties.
To prove Theorem \ref{6754387}, we will apply the so-called strong Laurent property for the Somos-5 equation formulated in \cite{Hone2}.

\begin{remark}
Theorem \ref{67543} was first proved in the work of Fomin and Zelevinsky \cite{Fomin}. More precisely, the Laurent property was proved for an infinite family of
recurrence relations of the form
\[
x_{n}x_{n+N}=\alpha x_{n+a}x_{n+N-a}+\beta x_{n+c}x_{n+N-c},
\]
known in the literature as Gale--Robinson equations \cite{Gale}. Here
\(N\geq 4\) and \(1\leq a<c\leq \lfloor N/2 \rfloor\). The proof
was based on an essential connection of these relations with the theory of
cluster algebras. 
\end{remark}

The article is organized as follows. In the next section we prove a theorem stating that each of the components of the solution \(\left(x_n, y_n\right)\) of system \eqref{45777}
satisfies the Somos-5 equation with the same pair of parameters \(\left(\alpha_0, \alpha_1\right)\).
This fact underlies the proof of the main theorem on the Laurent property. The proof of Theorem \ref{6754387} essentially relies on the theorem on the strong Laurent property for the Somos-5 equation. We formulate and discuss this theorem in Section \ref{67428766}. Section \ref{67420888} is devoted to the proof of Theorem \ref{6754387}. In Section \ref{5643987} we prove a theorem on the strong Laurent property for the system \eqref{45777} under study. This theorem can be used to construct integer sequences \(\left(x_n, y_n\right)\) when \(x_2, y_2\neq \pm 1\). In Section \ref{65320999} an example of such a sequence is given. Section \ref{8943298} is devoted to concluding remarks and discussion of possible practical applications of the sequences generated by the system of discrete equations studied in the article.

\section{Connection of the studied system of discrete equations with the Somos-5 equation}
\label{8742988}

In this section we present auxiliary lemmas establishing
the connection of system \eqref{45777} with the Somos-5 equation, and formulate
the corresponding theorem.

Let a non-singular solution \((x_n, y_n)\) of system \eqref{45777} be given,
determined by some set of data. Define the sequences
\begin{equation}
v_n  = \frac{x_n y_{n+1}}{y_n x_{n+1}},\; u_n  = \frac{y_n x_{n+1}}{x_n y_{n+1}}.
\label{45700077}
\end{equation}
\begin{lemma}
By virtue of system \eqref{45777} and \eqref{45700077} the pair of relations holds
\begin{equation}
\begin{aligned}
v_{n+2}\left(d+c\,v_{n+1}\right)=u_n\left(a+b\,u_{n+1}\right), \\
u_{n+2}\left(a+b\,u_{n+1}\right)=v_n\left(d+c\,v_{n+1}\right).
\end{aligned}
\label{457700007}
\end{equation}
\end{lemma}
\begin{proof}
The proof is carried out by direct verification.
\end{proof}
\begin{remark}
By virtue of \eqref{45700077} the following holds:
\begin{equation}
v_nu_n=1.
\label{67543000}
\end{equation}
In other words, this system can be reduced to an equation with respect to one function \(v_n\) or \(u_n\). If we consider this system
as an independent object, we can notice that \(p_n=v_nu_n\) is a 2-invariant of this system. That is, the relation
\(p_{n+2}=p_n\) holds. This is consistent with condition \eqref{67543000}, considered as a special case \(p_0=p_1=1\) for this 2-invariant.
\end{remark}
\begin{remark}
It should be noted that the conditions \(u_n=-c/d\) or \(v_n=-b/a,\) under which indeterminacies appear in system \eqref{457700007}, cannot hold for a non-singular solution of system \eqref{45777}. In other words, by requiring \(x_n\neq 0\) and \(y_n\neq 0\), we exclude these indeterminacies in system \eqref{457700007}.
\end{remark}

\begin{lemma}
System \eqref{457700007} with condition \eqref{67543000} admits
the invariant
\begin{equation}
H= a(u_n+u_{n+1})+d(v_n+v_{n+1})+b\,u_nu_{n+1}+c\,v_nv_{n+1}.
\label{89535555}
\end{equation}
\end{lemma}
\begin{proof}
Denote the left-hand side of \eqref{89535555} by \(H_n\). Then
\[
\begin{aligned}
H_{n+1}-H_n
&=a(u_{n+2}-u_n)+d(v_{n+2}-v_n)  \\
&\quad+b\,u_{n+1}(u_{n+2}-u_n)+c\,v_{n+1}(v_{n+2}-v_n)    \\
&=(u_{n+2}-u_n)(a+b\,u_{n+1})
 +(v_{n+2}-v_n)(d+c\,v_{n+1}).
\end{aligned}
\]
Substituting \(v_{n+2}\) and \(u_{n+2}\), by virtue of system \eqref{457700007}, into the last relation, we obtain \(H_{n+1}=H_n\). The lemma is proved.
\end{proof}

Note that the recurrence relation for the sequence
\((v_n)\) can be written in the form
\begin{equation}
v_nv_{n+1}v_{n+2}=\mathcal M(v_{n+1}),
\label{8954435}
\end{equation}
where
\(
\mathcal M(v)=\left(av+b\right)/\left(cv+d\right).
\)
The function \(\mathcal M\) defines a fractional-linear (M\"obius)
transformation \(v\mapsto \mathcal M(v)\) with coefficient matrix
\begin{equation}
M=\begin{pmatrix}a&b\\c&d\end{pmatrix}.
\label{58543200}
\end{equation}

As a result, we obtain a dynamical system which is the result of an infinite number of iterations of a birational map of recurrence type:
\[
\left(
\begin{array}{c}
\tilde v_0 \\
\tilde v_1
\end{array}
\right)
=\left(
\begin{array}{c}
v_1 \\[0.2cm]
\dfrac{\mathcal M (v_1)}{v_0v_1}
\end{array}
\right).
\]
The inverse map, as is easy to verify, is defined as
\[
\left(
\begin{array}{c}
 v_0 \\
 v_1
\end{array}
\right)
=\left(
\begin{array}{c}
\dfrac{\mathcal M\left(\tilde v_0 \right)}{\tilde v_0 \tilde  v_1}
 \\[0.3cm]
\tilde v_0
\end{array}
\right).
\]
Note that the inverse matrix \(M^{-1}\) does not participate here and therefore equation \eqref{8954435} is defined also for a degenerate matrix.
\begin{lemma}
System of equations \eqref{45777} admits the invariant
\begin{align}
H&= \left(\frac{y_{n}x_{n+1}}{x_{n}y_{n+1}}
+\frac{y_{n+1}x_{n+2}}{x_{n+1}y_{n+2}}\right)a
+\left(\frac{x_{n}y_{n+1}}{y_{n}x_{n+1}}
+\frac{x_{n+1}y_{n+2}}{y_{n+1}x_{n+2}}\right)d \notag \\
&\quad +\frac{y_nx_{n+2}}{x_ny_{n+2}} b
+\frac{x_ny_{n+2}}{y_nx_{n+2}} c,
\label{89765}
\end{align}
where \((x_n, y_n)\) is an arbitrary non-singular solution.
\end{lemma}

Define
\begin{equation}
V_n =v_n\left(d+c  v_{n+1}\right),\; U_n =u_n\left(a+b u_{n+1}\right).
\label{4549994328}
\end{equation}
\begin{lemma}
\label{497643}
By virtue of \eqref{457700007} and \eqref{4549994328}, the sequences \((V_n)\) and \((U_n)\)
satisfy the recurrence relations
\begin{equation}
V_nV_{n+1}V_{n+2}=\alpha_0+\alpha_1 V_{n+1},\:\; U_nU_{n+1}U_{n+2}=\alpha_0+\alpha_1 U_{n+1}
\label{466628}
\end{equation}
with parameters
\begin{equation}
\alpha_1=ad-bc,\;\; \alpha_0=a^2 c + d^2 b + b\,c\,H,
\label{4544328}
\end{equation}
where \(H\) is the invariant of equation \eqref{8954435}, determined by expression \eqref{89535555}.
\end{lemma}
As we can see, the parameter \(\alpha_1\) is expressed as the determinant of the coefficient matrix \eqref{58543200}. The parameter \(\alpha_0\), unlike \(\alpha_1\), depends on the initial data of the solution.
\begin{lemma}
\label{477777643}
For an arbitrary non-singular solution \((x_n, y_n)\) of system
\eqref{45777} the following relations hold
\begin{equation}
V_n=\frac{x_nx_{n+3}}{x_{n+1}x_{n+2}},\;\; U_n=\frac{y_ny_{n+3}}{y_{n+1}y_{n+2}}.
\label{458}
\end{equation}
\end{lemma}
\begin{proof}
These relations are proved by direct verification. In doing so, one needs to use the equations of system \eqref{45777} to express \(x_{n+3}\) and \(y_{n+3}\), the substitutions \eqref{45700077} and \eqref{4549994328}.
\end{proof}

As a result of this section, we obtain the main statement of this section --- a theorem following from Lemma \ref{497643} and Lemma \ref{477777643}.
\begin{theorem}
\label{564389}
By virtue of system \eqref{45777} the sequences \((x_n)\) and \((y_n)\)
satisfy the Somos-5 equation with the same parameters \(\alpha_0,\alpha_1\),
determined by formulas \eqref{4544328}. The quantity \(H\), given by
formula \eqref{89765}, is an invariant of system \eqref{45777}.
\end{theorem}
\begin{proof}
The proof consists of direct verification. Substituting the expressions
\eqref{458} for \(V_n\) and \(U_n\) into equations \eqref{466628},
written for each of these sequences, we obtain
the required result.
\end{proof}

\begin{remark}
It is well known (see, for example, \cite{Hone2}) that equation \eqref{466628} together with the substitution \eqref{458} leads to
the Somos-5 equation. However, in the present paper the substitutions \eqref{458} for \(V_n\) and \(U_n\) arise in a non-trivial way as a consequence of
system \eqref{45777} (see Lemma \ref{477777643}), and are not postulated in advance.
\end{remark}

\section{Connection between the invariants of the studied system and the Somos-5 equation}

It is known that the Somos-5 equation \eqref{408} admits the invariant
\cite{Hone2}
\begin{equation}
H_x=\frac{x_nx_{n+3}}{x_{n+1}x_{n+2}}
+\frac{x_{n+1}x_{n+4}}{x_{n+2}x_{n+3}}
+\alpha_1\left(\frac{x_{n+1}x_{n+2}}{x_{n}x_{n+3}}
+\frac{x_{n+2}x_{n+3}}{x_{n+1}x_{n+4}}\right)
+\alpha_0\frac{x^2_{n+2}}{x_{n}x_{n+4}}.
\label{67536}
\end{equation}
The analogous expression for the invariant \(H_y\) is obtained by replacing \(x_n\) with \(y_n\). Note that by virtue of the first substitution in \eqref{458} the expression for
\(H_x\) is rewritten as
\begin{equation}
H_x=V_n+V_{n+1}
+\alpha_1\left(\frac1{V_n}+\frac1{V_{n+1}}\right)
+\frac{\alpha_0}{V_nV_{n+1}}.
\label{677777636}
\end{equation}
Similarly, by virtue of the second substitution in \eqref{458}, we obtain
\begin{equation}
H_y=U_n+U_{n+1}
+\alpha_1\left(\frac1{U_n}+\frac1{U_{n+1}}\right)
+\frac{\alpha_0}{U_nU_{n+1}}.
\label{675777636}
\end{equation}
The following lemma can be regarded as a supplement to Theorem \ref{564389}.
\begin{lemma}
\label{76309}
For any non-singular solution of system \eqref{45777}, the value of the invariant \(H\), determined by \eqref{89535555}, coincides with the value of \(H_x\) (\(H_y\)).
\end{lemma}
\begin{proof}
We use the first relation in \eqref{466628} to eliminate \(\alpha_0\) in \eqref{677777636}. Then we use the first substitution in \eqref{4549994328} and express \(\alpha_1\) by virtue of \eqref{4544328}.
As a result, we obtain
\begin{align}
H_x&=V_n+V_{n+1}+V_{n+2}+\frac{\alpha_1}{V_{n+1}} \notag \\
&=\left(d\,v_n+c\,v_nv_{n+1}\right) +\left(d\,v_{n+1}+c\,v_{n+1}v_{n+2}\right) +\left(d\,v_{n+2}+c\,v_{n+2}v_{n+3}\right) \notag \\
&\quad +\dfrac{ad-bc}{d\,v_{n+1}+c\,v_{n+1}v_{n+2}}  \notag
\end{align}
Here one needs to eliminate \(v_{n+2}\) and \(v_{n+3}\), by virtue of the equation \(v_nv_{n+1}v_{n+2}=\mathcal M(v_{n+1})\) and verify that indeed \(H_x=H\), where \(H\) is determined by formula \eqref{89535555}. The calculations, although cumbersome, are algorithmic.
For \(H_y\) the reasoning is completely analogous with \(V_n\) replaced by \(U_n\). Hence, in particular, we obtain that \(H_x=H_y\).
\end{proof}

Thus, the sequences \((x_n)\) and \((y_n)\), which by virtue of Theorem \ref{564389} satisfy the same Somos-5 equation, have the same value of the invariant.
\begin{remark}
Note that according to Theorem~\ref{564389} on the connection of the studied system
with the Somos-5 equation and the work of Hone~\cite{Hone1}, the general solution
of the studied system~\eqref{45777} must be expressed through the
\(\sigma\)-function of Weierstrass associated with some
elliptic curve and its Jacobian. Formulas~\eqref{4544328}
uniquely determine from the data~\eqref{7666543} of the studied system
the coefficients of the corresponding Somos-5 equation. The Lemma~\ref{76309} proved in this
section says that the invariant~\eqref{67536}
of the Somos-5 equation can be replaced by the invariant~\eqref{89765}
(for \(n=0\)). By the formulas of Theorem~2.7 of \cite{Hone1}
the equation of the elliptic curve \(Y^2=4X^3-g_2X-g_3\) is determined.
The condition \(g_2^3-27g_3^2\neq 0\) (non-vanishing of the discriminant
of the curve) guarantees the non-singularity of this elliptic curve and, thereby, the correctness of the associated non-degenerate period lattice
and the function \(\sigma(z; g_2, g_3)\). In fact, one needs to understand how the constants appearing in the formulas for these solutions must be related.
A complete study of these questions
is beyond the scope of this article.
\end{remark}

\section{Strong Laurent property of the Somos-5 equation}
\label{67428766}

In principle, the idea of proving Theorem \ref{6754387} on the Laurent
property for system \eqref{45777} is as follows. First we
use Theorem \ref{564389}, according to which each of the
sequences \((x_n)\) and \((y_n)\) satisfies the Somos-5 equation.
Then we apply the Laurent property to this equation
(Theorem \ref{67543}) and transfer it to the original system. However,
such a course of action encounters technical difficulties.

It turns out that to prove Theorem \ref{6754387} it is more convenient
(and, essentially, necessary) to use the so-called strong
Laurent property for the Somos-5 equation, which is formulated with
the help of the following theorem proved in \cite{Hone2}. For
convenience we formulate this theorem in our notation.
\begin{theorem}
\label{453298}
By virtue of the Somos-5 equation \eqref{408}, the following holds:
\[
x_n\in \mathbb Z[x_0^{\pm 1}, x_1^{\pm 1}, x_2, x_3, x_4,
\alpha_0, \alpha_1, J_x],\;\;
n\ge 1,
\]
where \(J_x=\alpha_0+\alpha_1 H_x\).
\end{theorem}
Here \(H_x\) is expressed by \eqref{67536}. Recall that this is an invariant of the Somos-5 equation. Therefore \(J_x\) is also an invariant.
In the next section, for convenience, we will use the short notation
\(
\widetilde {\mathcal R}=\mathbb Z[x_0^{\pm 1}, x_1^{\pm 1}, x_2, x_3, x_4,
\alpha_0, \alpha_1, J_x].
\)

To illustrate this theorem we show three Laurent polynomials generated by the Somos-5 equation:
\begin{equation}
x_5= \frac{\alpha_1x_1x_4}{x_0}+\frac{\alpha_0x_2 x_3}{x_0},\;\;
x_6= \frac{\alpha_1^2 x_2 x_4}{x_0}
+\frac{\alpha_0\alpha_1x_2^2x_3}{x_0x_1}
+\frac{\alpha_0 x_3 x_4}{x_1},
\label{6539877}
\end{equation}
\begin{equation}
x_7=\frac{\alpha_0 x_3x_4}{x_0}  J_x
+\frac{\alpha_1^3 x_3 x_4}{x_0}
-\frac{\alpha_0\alpha_1^2 x_1 x_2 x_4}{x_0^2}
-\frac{\alpha_0^2\alpha_1 x_2^2 x_3}{x_0^2}.
\label{6539800077}
\end{equation}
\begin{remark}
In reality Theorem \ref{453298} does not strengthen Theorem \ref{67543}, but rather refines it. It says that negative powers in \(x_2,x_3,x_4\)
in all iterations \(x_n\) can be accounted for by a polynomial dependence on the indicated invariant \(J_x\) of the Somos-5 equation.
\end{remark}
\begin{remark}
It is important for us that Theorem \ref{453298} is also valid for \(n\le 0\). This fact can be obtained if we apply this theorem to the reversed sequence
\(
\tilde x_n=x_{5-n}.
\)
In other words, one needs to use the symmetry of the Somos-5 equation. Therefore the given formulation of the theorem actually covers all \(n\in\mathbb Z\).
\end{remark}
\begin{remark}
Theorem \ref{453298} is a strong statement, but it does not
lead to an algorithm for constructing integer
sequences generated by Somos-5. Indeed, setting
\(x_0,x_1=\pm1\) and \((x_2,x_3,x_4;\alpha_0,\alpha_1)\in\mathbb Z^5\),
we cannot be sure that the value of the invariant \(J_x\)
is integer. Note that, by virtue of the definition of \(J_x\), here in fact we need the
integrality of the product \(\alpha_1H_x\).
\end{remark}

Let us explain why it is necessary to use Theorem \ref{453298} to prove the main theorem. Recall that Theorem \ref{67543} states
that by virtue of the Somos-5 equation
\(
x_n\in \hat{\mathcal R}: =
\mathbb Z[x_0^{\pm1},x_1^{\pm1},x_2^{\pm1},x_3^{\pm1},x_4^{\pm1};
\alpha_0,\alpha_1]\;\;\;
\forall n\in\mathbb Z.
\)
Take an arbitrary polynomial \(\hat P\in \hat{\mathcal R}\). Suppose that \(x_3\) and \(x_4\) are iterations by virtue of system \eqref{45777}. Let us write out the expressions for these iterations explicitly:
\begin{equation}
x_3=d\frac{x_2y_1}{y_0}+c\frac{x_1y_2}{y_0},\;\; x_4=d^2\frac{x_2y_2}{y_0}+cd\frac{x_1y_2^2}{y_0y_1}+ac\frac{x_1x_2y_2}{x_0y_1}+bc\frac{x_2^2}{x_0}.
\label{786543}
\end{equation}
The problem is that negative powers of these iterations do not belong to the ring
\(
\mathcal R:=\mathbb Z[x_0^{\pm1},x_1^{\pm1},x_2^{\pm1},
y_0^{\pm1},y_1^{\pm1},y_2^{\pm1};a,b,c,d],
\)
but are general rational functions. Therefore the polynomial \(\hat P\), upon substitution of negative powers of \(x_3\) and \(x_4\), does not transform into an element of the ring \(\mathcal R\).
On the other hand, polynomials belonging to the ring \(\widetilde{\mathcal R}\) do not contain \(x_3\) and \(x_4\) in negative powers.

\section{Proof of Theorem \ref{6754387}}
\label{67420888}

Recall that this theorem states that all iterations generated by system (\ref{45777}) belong to the ring \(\mathcal R\).
For the iterations \(x_3\) and \(x_4\), determined by the expressions in \eqref{786543}, this is obvious.
By virtue of Lemma \ref{76309}, we can make the replacement \(H_x\rightarrow H\) or, in other words, the invariant of the Somos-5 equation can be replaced by the invariant of system (\ref{45777}). Taking this replacement into account, we obtain:
\(
\alpha_1=ad-bc\in\mathcal R,
\)
and
\(
\alpha_0=a^2 c + d^2 b + bc H\in\mathcal R.
\)
Moreover
\(
J=a^2 c + d^2 b+ a d H\in\mathcal R,
\)
since
\begin{equation}
H= \left(\frac{y_{0}x_{1}}{x_{0}y_{1}}
+\frac{y_{1}x_{2}}{x_{1}y_{2}}\right)a
+\left(\frac{x_{0}y_{1}}{y_{0}x_{1}}
+\frac{x_{1}y_{2}}{y_{1}x_{2}}\right)d
+\frac{y_0x_{2}}{x_0y_{2}} b+ \frac{x_0y_{2}}{y_0x_{2}} c \in\mathcal R.
\label{7865000043}
\end{equation}
Substituting
\(
x_3, x_4, \alpha_0, \alpha_1, J
\)
into an arbitrary Laurent polynomial \(\widetilde P\in \widetilde{\mathcal R}\), we obtain a mapping \(\widetilde{\mathcal R}\mapsto\mathcal R\). In particular, since any iteration \(x_n\) generated by the Somos-5 equation belongs to \(\widetilde {\mathcal R}\), then, taking these substitutions into account, it will belong to the ring \(\mathcal R\).
Similarly, by virtue of the symmetry of system (\ref{45777}), we obtain
\(
y_n\in\mathcal R\;\; \forall n\in\mathbb Z.
\)
Thus, Theorem \ref{6754387} is completely proved.

\section{Analogue of Theorem \ref{453298} for the studied system}
\label{5643987}

It turns out that, using Theorem \ref{453298}, one can prove an analogous theorem for the studied system \eqref{45777}.
\begin{theorem}
\label{6534766}
By virtue of the system of equations \eqref{45777}:
\[
x_n,\;y_n\in \mathcal R^{*}:=\mathbb Z[x_0^{\pm1},x_1^{\pm1},x_2,\;\; y_0^{\pm1},y_1^{\pm1},y_2;\,a, b, c, d, \xi, \eta]
\;\; \forall n\in\mathbb Z,
\]
where \(\xi=a\,d\,H\) and \(\eta=b\,c\,H\).
\end{theorem}
\begin{proof}
Recall that by Theorem \ref{453298}, by virtue of the Somos-5 equation, we have
\(
x_n\in\widetilde{\mathcal R}=\mathbb Z[x_0^{\pm1},x_1^{\pm1},x_2,x_3,x_4,\alpha_0,\alpha_1,J_x],
\)
for all \(n\in\mathbb Z\), where \(J_x=\alpha_0+\alpha_1H_x\). Here \(\alpha_0\) and \(\alpha_1\) are the parameters of the Somos-5 equation; \(H_x\) is the invariant of this equation.

Again suppose that \(x_3\) and \(x_4\) are iterations by virtue of system \eqref{45777}.
Recall that in this case, by virtue of Lemma \ref{76309}, we can make the replacement \(H_x\rightarrow H\), where \(H\) is the invariant of this system, determined by \eqref{7865000043}.

Note that \(x_3\) and \(x_4\), determined by \eqref{786543}, belong to \(\mathcal R^{*}\). Obviously,
\(
\alpha_1=ad-bc\in\mathcal R^{*},
\)
We use Lemma \ref{76309}, which gives us the right to make the replacement \(H_x\rightarrow H\) and correspondingly \(J_x\rightarrow J\). In this case we obtain the following:
\[
\begin{aligned}
J&=\alpha_0+\alpha_1H \\
&=a^2c+d^2b+bcH+(ad-bc)H \\
&=a^2c+d^2b+adH \\
&=a^2c+d^2b+\xi\in\mathcal R^{*}.
\end{aligned}
\]
and
\[
\begin{aligned}
\alpha_0&=a^2c+d^2b+bcH  \\
&=a^2c+d^2b+\eta\in\mathcal R^{*},
\end{aligned}
\]
Thus a mapping \(\widetilde{\mathcal R}\mapsto \mathcal R^{*}\) is defined, where \(\widetilde{\mathcal R}\) is the ring of Laurent polynomials in Theorem \ref{453298}.
The theorem is proved.
\end{proof}
\begin{remark}
Note that, just as in the case of Theorem \ref{453298}, this theorem essentially does not strengthen the ordinary Laurent property, but only refines
it. Speaking informally, all negative powers in the variables \(x_2\) and \(y_2\) are "hidden" here in the expressions for \(\xi\) and \(\eta\).
\end{remark}

\begin{remark}
As we have already said, Theorem \ref{453298} does not lead to an algorithm for constructing integer sequences for the Somos-5 equation. The same can be said about Theorem \ref{6534766}. It gives a sufficient criterion for integrality, but does not lead to an algorithm, since by choosing the conditions \(x_j, y_j=\pm 1\) for \(j=0, 1\) and \(\left(x_2, y_2; a, b, c, d\right)\in\mathbb Z^6\) we cannot be sure that \(\xi\) and \(\eta\), under these conditions, simultaneously take integer values. Nevertheless, one can find such an example.
\end{remark}

\section{Example of a pair of integer sequences}
\label{65320999}

Consider an example illustrating Theorem \ref{6534766}. Let \(x_0, x_1, y_0, y_1=1\) and
\(
\left(x_2, y_2, a, b, c, d\right)=\left(2, 3, 4, 2, 1, 4\right).
\)
Since \(x_2, y_2\neq \pm 1\), Theorem \ref{6754387} in this case does not guarantee the integrality of the sequences \(\left(x_n\right)\) and \(\left(y_n\right)\). This requires an additional condition, which is given by Theorem \ref{6534766}.
Note that although \(H=39/2\) is fractional, the quantities \(\xi=312\) and \(\eta=39\) are integer. Under these conditions Theorem \ref{6534766} guarantees the integrality of these sequences. Moreover, since we have taken all data positive and non-zero, this guarantees the non-singularity of \(\left(x_n\right)\) and \(\left(y_n\right)\).

In the following table the first nine terms of these sequences are shown.
\[
\begin{array}{c|ccccccccc}
n & 0 & 1 & 2 & 3 & 4 & 5 & 6 & 7 & 8\\ \hline
x_n & 1 & 1 & 2 & 11 & 164 & 4210 & 274828 & 51\,195\,896 & 19\,837\,002\,616\\
y_n & 1 & 1 & 3 & 16 & 194 & 6892 & 559\,512 & 80\,551\,288 & 34\,641\,433\,216
\end{array}
\]
Let us look at this example to see what the greatest common divisors of the numbers \(x_n\) and \(y_n\) are. Denote \(g_n=\gcd\left(x_n, y_n\right)\). The first nine terms of this sequence are as follows:
\[
\begin{array}{c|ccccccccc}
n & 0 & 1 & 2 & 3 & 4 & 5 & 6 & 7 & 8\\ \hline
g_n & 1 & 1 & 1 & 1 & 2 & 2 & 4 & 8 & 8
\end{array}
\]
Numerical computations for this example allow us to conjecture that \(g_n=2^{\varepsilon(n)}\), where \(\varepsilon(n)\) is a non-decreasing
sequence of non-negative integers. Note that in this case \(\det M=2\cdot7\). In other words, within the computations performed
the common divisors of \(x_n\) and \(y_n\) always turn out to be powers of two --- one of the prime divisors of the determinant of the coefficient matrix.

\section{Conclusion}
\label{8943298}

The main results of the article are three theorems. Theorem
\ref{564389} establishes that system \eqref{45777} generates two
sequences \((x_n)\) and \((y_n)\), each of which
satisfies the same Somos-5 equation with coefficients
\(\alpha_0\) and \(\alpha_1\), determined from the data of the system
by formulas \eqref{4544328}. Relying on Theorem \ref{453298} on the
strong Laurent property for the Somos-5 equation and Theorem
\ref{564389}, we proved the main Theorem \ref{6754387} on the Laurent
property for system \eqref{45777}. Finally, Theorem \ref{6534766}
gives an analogue of the strong Laurent property for system \eqref{45777}.

One can pose the following inverse problem. Let some solution \((x_n)\) of the Somos-5 equation with fixed coefficients \(\alpha_0\) and \(\alpha_1\) be given. It is required to find a solution \((y_n)\) of the same Somos-5 equation such that the pair \((x_n, y_n)\) is a solution of system \eqref{45777} with some coefficients.

Note that formally such a solution always exists. Indeed, take \(y_n=x_{n+2}\). It is easy to verify that \((x_n, x_{n+2})\) is a solution of system \eqref{45777} with \(a=\alpha_1\), \(b=\alpha_0\), \(c=0\), \(d=1\). However, it is not interesting and we fundamentally do not consider it as a solution. The search for non-trivial solutions of this inverse problem leads to a system
of polynomial equations with respect to the data of system \eqref{45777}.
As computations show, a non-trivial solution may not
exist, and if it exists, it may not be unique.

Let us turn to the question of the potential possibilities of practical
application of the sequences (not necessarily integer)
generated by system \eqref{45777}. First of all, note that this
system is quite simple in construction and therefore may be
attractive from a practical point of view. Since both components
of the solution satisfy the Somos-5 equation, it is natural to start with
applications of the sequences generated by this equation, although
it is worth noting that there exists a whole class of bilinear
(quadratic) recurrence relations with similar properties.
As an example, one can take the Gale--Robinson equations mentioned above.

Let us list some specific features that are interesting from the point of
view of application in cryptography.

The sequences generated by the Somos-5 equation are closely related to elliptic curves \cite{Hone2} ---
mathematical objects on which modern cryptography is based. This allows one to construct cryptographic algorithms
similar to existing ones, but relying on the properties of Somos sequences \cite{Illarionov}.

The second feature is the finiteness of the rank of the sequences generated by the recurrence relations given in the article (see, for example, \cite{Ustinov1,Ustinov}).
Note that among the bilinear equations given there are both equations for which the finiteness of rank is proved (for example, Somos-5) and equations for which this question remains open. For sequences of finite rank
a multitude of identities holds, allowing one to compute iterations \(x_n\) rapidly for large \(n\).

The third feature is special arithmetic properties. The study of the arithmetic properties of integer sequences generated by Somos-5 was
initiated in \cite{Robinson} and then continued, for example, in \cite{Ustinov,Kamp}.

Apparently, pairs of non-singular integer sequences
generated by system \eqref{45777} possess specific
arithmetic properties that may find application. However,
their study is beyond the scope of this article.

\section*{Acknowledgements}

This work was carried out within the framework of  the state assignment of the Ministry of Education
and Science of the Russian Federation on the project No.  126021217175-3.

\end{document}